\documentclass[%
 reprint,
 amsmath,amssymb,
 prl,
 floatfix,
 groupedaddress,
]{revtex4-2}

\usepackage{graphicx}
\usepackage{dcolumn}
\usepackage{multirow}
\usepackage{bm}
\usepackage{bbold}
\usepackage{braket}

\begin{document}


\title{\textit{Ab initio} \textit{GW}-BSE theory of optical activity in $\alpha$-quartz}

\author{Xiaoming Wang}
\email{xiaoming.wang@utoledo.edu}
\affiliation{%
 Department of Physics and Astronomy, Wright Center for Photovoltaics Innovation and Commercialization, The University of Toledo, Toledo, Ohio 43606, USA
}%

\author{Yanfa Yan}%
\email{yanfa.yan@utoledo.edu}
\affiliation{%
 Department of Physics and Astronomy, Wright Center for Photovoltaics Innovation and Commercialization, The University of Toledo, Toledo, Ohio 43606, USA
}%


\begin{abstract}
We present an \textit{ab initio} many-body theory of optical activity in solids within the \textit{GW}-BSE framework. Dielectric spatial dispersion is formulated in two complementary forms: exciton envelope modulation and sum-over-exciton-states expansion. Our application to $\alpha$-quartz reveals that the envelope-modulated formulation captures the low-frequency region, whereas the sum-over-exciton-states formulation is essential to reproduce the correct full frequency dependence. Comparisons with the independent-particle approximation and simple local-field corrections further highlight the decisive role of excitonic many-body effects in shaping the spectral dispersion of optical activity in solids.
\end{abstract}

\maketitle

Optical activity refers to the different electromagnetic response of an optically active, most often chiral, material to right- and left-circularly polarized light. Its most familiar manifestation is optical rotation, in which the plane of polarization of linearly polarized light rotates as it propagates through the medium. First observed by Arago in $\alpha$-quartz ~\cite{arago1811,Lowry1922} and later widely recognized in chiral molecules~\cite{flack2009}, optical activity is now a key diagnostic and functional property in chemistry, biology, and materials science.

Despite more than two centuries since its discovery in $\alpha$-quartz, an \textit{ab initio} prediction of the full frequency optical activity in this prototypical chiral crystal remains elusive. By contrast, \textit{ab initio} molecular optical activity is well established through multipole expansions of charge and current distributions~\cite{Barron2004}. The lack of a comparable multipole theory for periodic solids originates from the ill-defined position operator $\mathbf{r}$ in periodic boundary conditions, which enters explicitly in the multipole moments.

From the perspective of crystal optics, optical activity originates microscopically from the spatial dispersion of the dielectric response \cite{Landau1984}. The first \textit{ab initio} formulations for crystalline solids appeared in the 1990s, when Zhong et al. \cite{Zhong1992,Zhong1993} expanded the dielectric tensor to first order in wavevector $\mathbf{q}$ within the independent-particle approximation (IPA). Applied to $\alpha$-quartz, the resulting optical rotation was found to be significantly underestimated. Subsequent work showed that local-field corrections (LFCs) are essential in $\alpha$-quartz~\cite{Jonsson1996}. When combined with a tuned scissors shift $\Delta$ applied to the single-particle gap, LFCs can reproduce the static-limit optical rotation, underscoring the crucial role of the band gap. A modern reformulation of spatial dispersion within IPA was later developed by Malashevich et al. \cite{Malashevich2010prb} and implemented using Wannier interpolation~\cite{Tsirkin2018}, yielding excellent agreement for materials such as trigonal Te. More recently, multipole theories of optical activity in solids within IPA have been introduced and implemented in state-of-the-art DFT codes \cite{Rerat2021,Balduf2022,multunas2023,pozo2023,wang2023}, and new LFC treatments have emerged based on orbital relaxation \cite{Desmarais2023} or density-functional perturbation theory \cite{Zabalo2023}. Across all recent studies on $\alpha$-quartz \cite{Jonsson1996,Desmarais2023,Zabalo2023}, two conclusions consistently emerge: optical activity is extremely sensitive to the crystal structure, and LFCs are required to obtain the correct order of magnitude. However, all existing calculations rely on the single-particle band structure, often with a functional-dependent or empirically tuned band gap, which limits predictive power. Moreover, previous works have focused exclusively on the static limit, while the full frequency dependence, or optical rotatory dispersion, has not been reported.

$\alpha$-quartz is a wide-gap insulator, and the state-of-the-art theory for its optical response is the Bethe–Salpeter equation (BSE) \cite{Rohlfing2000,Onida2002}, which explicitly incorporates electron–hole interactions and is built atop the many-body \textit{GW} quasiparticle framework \cite{Hedin1965,Hybertsen1986}. The \textit{GW}–BSE approach accurately reproduces the long-wavelength optical properties of $\alpha$-quartz \cite{Chang2000,kresse2012}, suggesting that its extension to finite $\mathbf{q}$ should yield a more reliable description of optical activity than IPA-based methods or with simple LFCs. Very recently, optical activity has been evaluated within \textit{GW}–BSE framework for chiral halide perovskites \cite{li2024,xu2025} with a minimal-coupling method that retains the position operator. In this Letter, we develop an \textit{ab initio} many-body theory of optical activity for crystalline solids by explicitly expanding the \textit{GW}–BSE dielectric function to first order in wavevector $\mathbf{q}$. Using exciton envelope modulation and a sum-over-exciton-states (SOXS) expansion, we derive two complementary formulations for optical activity and exciton multipoles. Applied to $\alpha$-quartz, our formulation yields optical rotatory dispersion in excellent agreement with experiment, resolving a long-standing challenge in the predictive description of optical activity in solids. 

We expand the dielectric tensor $\epsilon_{ij}(\omega, \mathbf{q})$ to first order in $\mathbf{q}$
\begin{equation}
    \epsilon_{ij}(\omega, \mathbf{q}) = \epsilon_{ij}(\omega, \mathbf{0}) + iq_l \gamma_{ijl}(\omega) + \mathcal{O}(\mathbf{q}^2),
\end{equation}
where $\gamma_{ijl}$ is the optical activity tensor. The experimentally measured optical rotation $\rho_{ijl} (\omega)$ and circular dichroism $\theta_{ijl} (\omega)$ are related to $\gamma_{ijl} (\omega)$ via 
\begin{equation}
    \rho_{ijl}(\omega) + i \theta_{ijl}(\omega) = \frac{\omega^2}{2c^2} \gamma_{ijl} (\omega) .
\end{equation} 
Including excitonic effects, the off-diagonal transverse dielectric function can be written as \cite{agranovich1984} (omitting the antiresonant term for clarity)
\begin{equation}
    \epsilon_{ij}(\omega,\mathbf{q}) = -\frac{4\pi e^2}{\Omega \omega^2 \hbar} \sum_\lambda \frac{\rho_{\lambda,i}^\ast (\mathbf{q}) \rho_{\lambda,j}(\mathbf{q})}{\omega - \omega_\lambda (\mathbf{q}) + i\eta} \, ,
\end{equation}
where $\Omega$ is the unit cell volume, $\lambda$ labels the exciton state, $\eta$ is a positive infinitesimal, and $i,j$ denote Cartesian components. The $\mathbf{q}$ dependence enters through the exciton dispersion $\omega_\lambda (\mathbf{q})$ and the oscillator strength $\bm{\rho}_\lambda (\mathbf{q})$. The contribution from energy dispersion is formally analogous to that of independent particle case~\cite{wang2023,pozo2023}. In particular, for 3D bulk materials, the linear term $\partial_\mathbf{q} \omega_\lambda(\mathbf{q})|_{\mathbf{q}=0}$ vanishes because the exciton Hamiltonian is quadratic in $\mathbf{q}$ near the zone center~\cite{qiu2021}. We therefore focus exclusively on the $\mathbf{q}$ dependence of the oscillator strength in this work.

The exciton oscillator strength is
\begin{equation}
    \bm{\rho}_\lambda (\mathbf{q})=\frac{1}{2}\langle \Psi_\lambda (\mathbf{q}) |\{e^{i\mathbf{q \cdot r}}, \mathbf{v}\}|0\rangle ,
\end{equation}
where the curly bracket indicates the anticommutator, $\mathbf{v}$ is the velocity operator characterizing the optical transition amplitude from the ground state $|0\rangle$ to the exciton state, which is expanded in the electron–hole basis as 
\begin{equation}
    |\Psi_\lambda (\mathbf{q}) \rangle = \sum_{cv\mathbf{k}} A_{cv\mathbf{k}}^\lambda(\mathbf{q}) |c\mathbf{k+q}\rangle_e \otimes |v\mathbf{k}\rangle_h .
\end{equation}
The exciton envelope $A_{cv\mathbf{k}}^\lambda (\mathbf{q})$, which weights the electron state $|c\mathbf{k+q}\rangle_e$ and hole state $|v\mathbf{k}\rangle_h$, is obtained by solving BSE:
\begin{equation}
    \mathcal{H}_{tt^\prime}^\mathrm{BSE}(\mathbf{q}) A_{t^\prime}^\lambda (\mathbf{q}) = \hbar \omega_\lambda (\mathbf{q}) A_{t}^\lambda (\mathbf{q}) ,
\end{equation}
with the Hamiltonian 
\begin{equation}
    \mathcal{H}_{tt^\prime}^\mathrm{BSE}(\mathbf{q}) = \hbar \omega_t (\mathbf{q}) \delta_{tt^\prime} + \langle t|\bar{v}-W|t^\prime \rangle(\mathbf{q}), 
\end{equation}
where $t \equiv cv\mathbf{k}$, $\bar{v}$ is the exchange interaction encoding local-field effects, and $W$ is the screened Coulomb interaction that accounts for electron–hole attraction. By neglecting $W$, we recover the IPA with a simple LFC, which we use to assess how local fields alone modify the optical activity.

A common treatment expresses the exciton oscillator strength as
\begin{equation}
    \bm{\rho}_\lambda (\mathbf{q})=\sum_{cv\mathbf{k}} A_{cv\mathbf{k}}^\lambda (\mathbf{q}) \mathbf{X}_{cv\mathbf{k}} (\mathbf{q}) \, ,
\end{equation}
where $\mathbf{X}_{cv\mathbf{k}}(\mathbf{q})=\frac{1}{2} \langle c\mathbf{k+q} |\{e^{i\mathbf{q \cdot r}}, \mathbf{v}\}|v\mathbf{k}\rangle$. Inserting a complete Bloch set $\sum_{n}|u_{n\mathbf{k}}\rangle \langle u_{n\mathbf{k}}| = 1$ between $e^{i\mathbf{q \cdot r}}$ and $\mathbf{v}$ and expanding for $\mathbf{q} \rightarrow \mathbf{0}$ to linear order in $\mathbf{q}$, one obtains (suppressing $\mathbf{k}$ for clarity)
\begin{equation}
    \mathbf{X}_{cv} (\mathbf{q})= \mathbf{V}_{cv} + \frac{i}{2} \sum_{n} \Big[(\mathbf{q} \cdot \bm{\mathcal{A}}_{cn}) \mathbf{V}_{nv} + \mathbf{V}_{cn} (\bm{\mathcal{A}}_{nv} \cdot \mathbf{q}) \Big],
\end{equation}
where $\bm{\mathcal{A}}_{nm} = \langle u_{n\mathbf{k}} | i\bm{\nabla}_\mathbf{k}| u_{m\mathbf{k}} \rangle$ and $\mathbf{V}_{nm} =  \langle u_{n\mathbf{k}} | \mathbf{v}_\mathbf{k}| u_{m\mathbf{k}} \rangle$ are the Berry connection and velocity matrix element, respectively. Defining the rank-2 tensor 
\begin{equation}
    \mathbf{W}_{cv}=\frac{1}{2} \sum_n (\bm{\mathcal{A}}_{cn} \otimes \mathbf{V}_{nv} + \mathbf{V}_{cn} \otimes \bm{\mathcal{A}}_{nv}),
\end{equation}
its antisymmetric part and symmetric part correspond to the electron magnetic dipole and electric quadrupole contributions, respectively: 
\begin{equation}
    \mathbf{M} = \frac{1}{2}(\mathbf{W} - \mathbf{W}^T), \quad \mathbf{Q} = \frac{1}{2} (\mathbf{W} + \mathbf{W}^T).
\end{equation}
This reproduces the previous multipole formulation~\cite{wang2023,pozo2023}, up to a different grouping of terms. We insert $\mathbf{M}$ and $\mathbf{Q}$ directly into the closed-form expression for the multipole theory~\cite{wang2023} to calculate the optical activity tensor. We note in passing that, within the IPA, the gauge-dependent intraband Berry connection terms appearing in Eq.~(9) cancel while evaluating the dielectric function~\cite{wang2023,pozo2023}. For exciton with envelope, however, an analogous cancellation is not guaranteed. We therefore neglect the intraband Berry connection terms. These intraband contributions are expected to be small in the absence of dense band degeneracies~\cite{xu2025}. 

The $\mathbf{q}$ dependence of the exciton envelope, $\partial_\mathbf{q} A_{cv\mathbf{k}}^\lambda(\mathbf{q})|_{\mathbf{q}=0}$, arises from the $\mathbf{q}$-dependent BSE Hamiltonian. For 3D bulk materials, this linear-in-$\mathbf{q}$ contribution can be shown to vanish using second-order perturbation theory thanks to the quadratic dependence of $\mathcal{H}^{\mathrm{BSE}}(\mathbf{q})$~\cite{qiu2021}. We therefore expand $\bm{\rho}_\lambda (\mathbf{q})$ in terms of exciton multipoles constructed from envelope-modulated Bloch multipoles as in Eq.~(8).  With these exciton multipoles, the optical activity is evaluated in direct analogy to the IPA multipole formulation~\cite{wang2023,pozo2023}. 

One caveat of this formulation is the Hamiltonian dependence of the velocity operator, $\mathbf{v} = (i/\hbar) [\mathcal{H}, \mathbf{r}]$. In principle, $\mathbf{v}$ in Eq.~(4) should be consistent with the excitonic Hamiltonian. In the Bloch-multipole formulation, however, a DFT or \textit{GW} single-particle Hamiltonian is typically employed. This inconsistency has been analyzed and corrected for the dielectric function in the long-wavelength limit~\cite{sangalli2017}. A direct remedy for the spatial-dispersion regime is less straightforward. We therefore retain both treatments and refer to them as the $\mathbf{v}_\mathrm{DFT}$ and $\mathbf{v}_\textit{GW}$ schemes.

\begin{table*}
    \caption{\label{tab:1} Optical rotation $\rho/(\hbar \omega)^2$ in deg/[mm (eV)\textsuperscript{2}] of $\alpha$-quartz in static limit. $^a$ Ref.~\cite{Zhong1993}. $^b$ Ref.~\cite{Zabalo2023}, sign adjusted for opposite handedness. $^c$ Ref.~\cite{Jonsson1996}, value for $\Delta=0$ eV (6.8) and $\Delta=1.8$ eV (5.6).}
    \begin{ruledtabular}
        \begin{tabular}{ccccccccc}
          \multirow{2}{*}{Method} & \multicolumn{2}{c}{\quad IPA} & \multicolumn{2}{c}{\qquad LFC} & \multicolumn{3}{c}{\textit{GW}-BSE} & \multirow{2}{*}{Exp.} \\
          & DFT & \textit{GW} & DFT & \textit{GW} & $\mathbf{v}_\mathrm{DFT}$ & $\mathbf{v}_{GW}$ & $\mathbf{v}_\mathrm{opt}$ \\
          \hline
          this work & -0.7 & -0.4 & 6.5 & 2.6 & 3.4 & 5.8 & 5.1 &  \\
          literature & 0.7~\footnotemark[1], -0.7~\footnotemark[2] & & -4.9~\footnotemark[2], 6.8~\footnotemark[3], 5.6~\footnotemark[3] & & & & & 4.6 $\pm$ 0.1~\footnotemark[3]
        \end{tabular}
    \end{ruledtabular}
\end{table*}

We apply our formulation to compute the optical rotatory dispersion $\rho (\omega)$ of $\alpha$-quartz (space group $P3_221$). We adopt the experimental lattice constants~\cite{lager1982} 
$a=4.914$ \AA\ and $c=5.406$ \AA. DFT, \textit{GW}, and BSE calculations are performed using \textsc{vasp}~\cite{Kresse1996cms,Kresse1996prb}. We employ the PBE functional~\cite{Perdew1996} for DFT. For \textit{GW}, we use the $GW_0$ scheme~\cite{shishkin2007}, include 1024 bands~\cite{supp_data}, and perform basis-size extrapolation~\cite{klime2014}. The BSE Hamiltonian is constructed using 18 valence and 24 conduction bands. A  $10\times10\times10$ $\Gamma$-centered k-mesh is used to sample the Brillouin zone~\cite{supp_data}. These settings are comparable to, or more stringent than, those used in previous studies~\cite{kresse2012}. The DFT direct gap at $\Gamma$ is 6.3 eV, which increases to 10.0 eV after \textit{GW} correction. The resulting optical gap (first exciton peak) from BSE is 8.9 eV. These values are consistent with prior calculations~\cite{Chang2000,kresse2012}. 

\begin{figure}[b]
\includegraphics[width=1.0\linewidth]{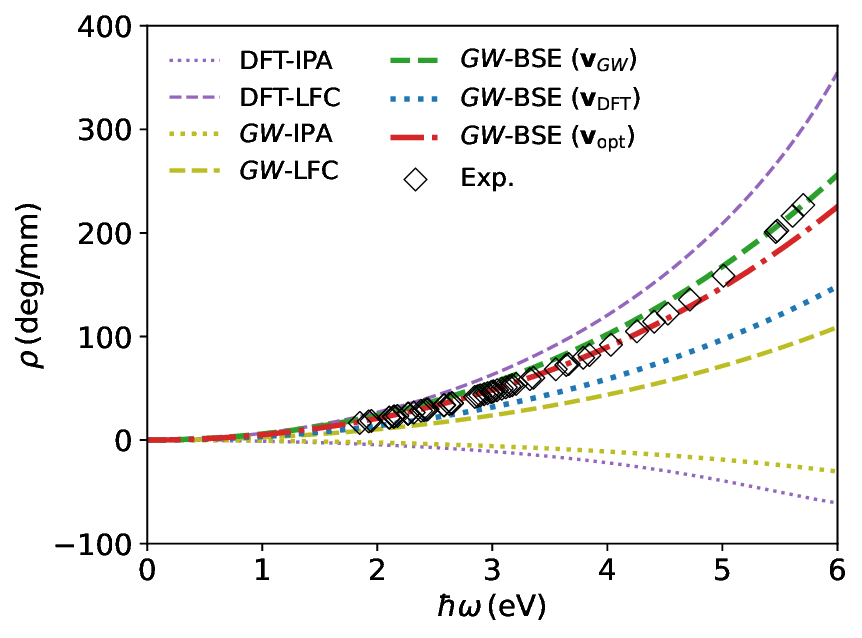}
\caption{\label{fig1} Optical rotatory dispersion of $\alpha$-quartz. Experimental points are taken from Ref.~\cite{lowry1964}.}
\end{figure} 

Fig.~\ref{fig1} shows the calculated optical rotatory dispersion for $\mathbf{q}$ along the optic axis ($\rho_{xyz}$ component) compared with experiment~\cite{lowry1964}. The optical rotation is often expressed as $\bar{\rho} = \rho/(\hbar \omega)^2$, which remains finite in the static limit and is listed in Table~\ref{tab:1}. A long-standing issue in the field is the sign inconsistency between different theoretical and experimental reports. In our case, the IPA result is negative, while the LFC and \textit{GW}-BSE results are positive. Ignoring the overall sign, our IPA value $\bar{\rho}=0.7$ agrees with previous calculations~\cite{Zhong1993,Zabalo2023}. With LFC on top of DFT, we obtain $\bar{\rho}=6.5$, in excellent agreement with the earlier value of 6.8~\cite{Jonsson1996}, and larger than the DFPT result of 4.9~\cite{Zabalo2023}. We note that optical activity is highly sensitive to the crystal structure, as also emphasized in Ref.~\cite{Zabalo2023,Desmarais2023}. Both our work and Ref.~\cite{Zhong1993} use the experimental structure, whereas Ref.~\cite{Zabalo2023} employs a LDA-relaxed structure, which likely accounts for part of the agreement and discrepancy among these results. In addition to structural sensitivities, optical activity is also strongly affected by the band gap. Since \textit{GW} significantly increases the DFT band gap, the corresponding IPA and LFC values are substantially reduced.

Within the \textit{GW}-BSE framework, the calculated static $\bar{\rho}$ values for the $\mathbf{v}_\mathrm{DFT}$ and $\mathbf{v}_{GW}$ schemes are 3.4 and 5.8, respectively, underestimating and overestimating the experimental value of 4.6. This trend is consistent with the fact that the DFT and \textit{GW} gaps under- and overestimate the experimental optical gap, respectively. To mitigate this, we introduce a scissors shift $\Delta$ applied to the DFT or \textit{GW} bands when evaluating the velocity matrix elements, chosen to match the BSE optical gap. We denote this as the $\mathbf{v}_\mathrm{opt}$ scheme. With this scheme, we obtain $\bar{\rho}=5.1$, in improved agreement with experiment. Beyond the static limit, the \textit{GW}-BSE-$\mathbf{v}_\mathrm{opt}$ method also reproduces the frequency dependence of $\rho (\omega)$ in the low-energy region ($< 5$ eV), with noticeable deviations emerging at higher energies. This breakdown reflects the fact that the scissors shift correctly aligns the lowest exciton with the electronic gap but leaves higher-lying exciton states unmatched with deeper bands, thereby neglecting dynamical electron-hole effects that become important for optical activity at higher frequencies.

To accurately treat the exciton oscillator strength in Eq.~(4), we insert a complete excitonic set, $\sum_\mu | \Psi_\mu (\mathbf{q}) \rangle \langle \Psi_\mu (\mathbf{q}) | = 1$, between $e^{i\mathbf{q \cdot r}}$ and $\mathbf{v}$, in direct analogy with the single-particle case. This yields
\begin{equation}
    \begin{split}
        \bm{\rho}_\lambda (\mathbf{q}) = \frac{1}{2} \sum_\mu \bigg [ & \langle \Psi_\lambda (\mathbf{q}) | e^{i\mathbf{q \cdot r}} | \Psi_\mu  \rangle \langle \Psi_\mu  | \mathbf{v} | 0 \rangle \\
        + & \langle \Psi_\lambda (\mathbf{q}) | \mathbf{v} | \Psi_\mu (\mathbf{q}) \rangle \langle \Psi_\mu (\mathbf{q}) | e^{i\mathbf{q \cdot r}} | 0 \rangle \bigg ],
    \end{split}
\end{equation}
where $\Psi_\mu \equiv \Psi_\mu (\mathbf{q=0})$. We can now use $\mathbf{v} = (i/\hbar)[\mathcal{H}^\mathrm{BSE}, \mathbf{r}]$ to ensure that the velocity operator is consistent with the excitonic Hamiltonian. 

There are four matrix elements in Eq.~(12). To evaluate the first one, we use the second-quantized exciton wave functions
\begin{equation}
\begin{split}
    \Psi_{\lambda}(\mathbf{q}) &= \sum_{cv\mathbf{k}} A_{cv\mathbf{k}}^\lambda (\mathbf{q}) c_{c\mathbf{k+q}}^\dagger c_{v\mathbf{k}} |0\rangle, \\
    \Psi_{\mu} &= \sum_{c' v'\mathbf{k}'} A_{c'v'\mathbf{k}'}^{\mu}  c_{c'\mathbf{k}'}^\dagger c_{v'\mathbf{k}'} |0\rangle
\end{split}
\end{equation}
(with $A_{cv\mathbf{k}} \equiv A_{cv\mathbf{k}}(\mathbf{q=0})$) and expand
\begin{equation}
    e^{i\mathbf{q \cdot r}} = \sum_{nn'\mathbf{k}''} \langle n\mathbf{k}''+\mathbf{q}|e^{i\mathbf{q \cdot r}}|n' \mathbf{k}'' \rangle c_{n\mathbf{k}''+\mathbf{q}}^\dagger c_{n' \mathbf{k}''} \, .
\end{equation}
After some lengthy algebra~\cite{supp_data}, we obtain
\begin{equation}
    \langle \Psi_\lambda (\mathbf{q}) | e^{i\mathbf{q \cdot r}} | \Psi_\mu  \rangle = \delta_{\lambda \mu} + i\mathbf{q} \cdot \bm{\mathcal{R}}_{\lambda \mu} + \mathcal{O}(\mathbf{q}^2),
\end{equation}
where the matrix element $\bm{\mathcal{R}}_{\lambda \mu}$ has \emph{interband} and \emph{intraband} contributions,
\begin{equation}
\begin{split}
    \bm{\mathcal{R}}_{\lambda \mu}^{\mathrm{inter}} = \sum_{cv\mathbf{k}} \bigg [ &\sum_{c' \neq c}(A_{cv\mathbf{k}}^{\lambda})^\ast  A_{c' v \mathbf{k}}^\mu \bm{\mathcal{A}}_{cc' \mathbf{k}} \\
    -&\sum_{v' \neq v} (A_{cv\mathbf{k}}^{\lambda})^\ast  A_{cv' \mathbf{k}}^\mu \bm{\mathcal{A}}_{v' v \mathbf{k}} \bigg ],
\end{split}
\end{equation}
\begin{equation}
    \bm{\mathcal{R}}_{\lambda \mu}^{\mathrm{intra}} = \sum_{cv\mathbf{k}} i (A_{cv\mathbf{k}}^{\lambda})^\ast \tilde{\partial}_\mathbf{k} A_{c v \mathbf{k}}^\mu \, ,
\end{equation}
with the covariant derivative 
\begin{equation}
    \tilde{\partial}_\mathbf{k} A_{c v \mathbf{k}} = \partial_\mathbf{k} A_{c v \mathbf{k}} - i (\bm{\mathcal{A}}_{cc\mathbf{k}} - \bm{\mathcal{A}}_{vv\mathbf{k}})A_{c v \mathbf{k}} \, .
\end{equation}
The \emph{intra} term encodes geometric corrections associated with the $\mathbf{k}$-space variation of the exciton envelope and strictly ensures gauge covariance. For a 3D trivial insulator such as $\alpha$-quartz, where the relevant excitons are Wannier-like and dominated by a small region of the Brillouin zone with smooth band geometry, $\bm{\mathcal{R}}_{\lambda \mu}^{\mathrm{intra}}$ is parametrically small compared to the interband contribution $\bm{\mathcal{R}}_{\lambda \mu}^{\mathrm{inter}}$. In this work we neglect $\bm{\mathcal{R}}_{\lambda \mu}^{\mathrm{intra}}$ and retain only $\bm{\mathcal{R}}_{\lambda \mu}^{\mathrm{inter}}$ in the optical-activity calculation. In materials with pronounced band geometry (e.g., large Berry curvature or strong quantum-metric effects, as often encountered in topological or near-topological systems), the intraband term can become significant and should be retained. It follows that $\bm{\mathcal{R}}_{\lambda \mu}$ is precisely the inter-exciton transition dipole moment. One can instead start from $\langle \Psi_\lambda |\mathbf{r}|\Psi_\mu \rangle$ using $\mathbf{r} = \sum_{nm\mathbf{k}''\mathbf{k'''}} \langle u_{n\mathbf{k}''}|\mathbf{r}|u_{m\mathbf{k}'''} \rangle c_{n\mathbf{k}''}^\dagger c_{m\mathbf{k}'''}$ and recover Eqs.~(16--17) by similar algebra~\cite{supp_data}. The quantity $\bm{\mathcal{R}}_{\lambda \mu}$ has been widely used in nonlinear optics~\cite{pedersen2015,taghizadeh2017,taghizadeh2018,ruan2024}.

The second matrix element of Eq.~(12) is straightforward: 
\begin{equation}
    \bm{\mathcal{V}}_\mu = \langle \Psi_\mu |\mathbf{v}|0 \rangle =i\omega_\mu \langle \Psi_\mu | \mathbf{r}|0\rangle= i\omega_\mu \bm{\mathcal{R}}_\mu .
\end{equation}
For the last matrix element, we find that the zeroth order vanishes, leaving
\begin{equation}
\begin{split}
    \langle \Psi_\mu (\mathbf{q}) | e^{i\mathbf{q \cdot r}} | 0 \rangle  &=\sum_{cv\mathbf{k}} A_{cv\mathbf{k}}^\mu (\mathbf{q}) \langle u_{c\mathbf{k+q}}| u_{v\mathbf{k}}\rangle \\
    &= i\mathbf{q} \cdot  \bm{\mathcal{R}}_\mu + \mathcal{O} (\mathbf{q}^2).
\end{split}
\end{equation}
Thus, to expand $\bm{\rho}_\lambda (\mathbf{q})$ to first-order in $\mathbf{q}$, the third matrix element in Eq.~(12) is taken at zeroth order, giving
\begin{equation}
    \bm{\mathcal{V}}_{\lambda \mu} = \langle \Psi_\lambda  | \mathbf{v} | \Psi_\mu  \rangle = i\omega_{\lambda \mu} \bm{\mathcal{R}}_{\lambda \mu}.
\end{equation}
Combining Eqs.~(12, 15, 19--21), we obtain the first-order expansion of the oscillator strength,
\begin{equation}
        \bm{\rho}_{\lambda} (\mathbf{q}) = \bm{\mathcal{V}}_\lambda + \frac{i}{2} \sum_{\mu}  \bigg [(\mathbf{q} \cdot \bm{\mathcal{R}}_{\lambda \mu} )\bm{\mathcal{V}}_\mu + \bm{\mathcal{V}}_{\lambda \mu} (\bm{\mathcal{R}}_\mu \cdot \mathbf{q}) \bigg ], 
\end{equation}
which has the same structure as Eq.~(9) for the single-particle case. In analogy, we define the tensor
\begin{equation}
    \bm{\mathcal{W}}_\lambda=\frac{1}{2}\sum_\mu ( \bm{\mathcal{R}}_{\lambda \mu} \otimes \bm{\mathcal{V}_\mu} + \bm{\mathcal{V}_{\lambda \mu}} \otimes \bm{\mathcal{R}}_\mu ),
\end{equation}
from which the exciton magnetic-dipole and electric-quadrupole contributions follow as
\begin{equation}
    \bm{\mathcal{M}} = \frac{1}{2} (\bm{\mathcal{W}}-\bm{\mathcal{W}}^T), \quad \bm{\mathcal{Q}} = \frac{1}{2} (\bm{\mathcal{W}}+\bm{\mathcal{W}}^T).
\end{equation}
We refer to this as the SOXS formulation, to distinguish it from the exciton envelope modulation approach. The SOXS formulation is fully consistent with the \textit{GW}-BSE framework and free of gauge ambiguities.

\begin{figure}
\includegraphics[width=0.9\linewidth]{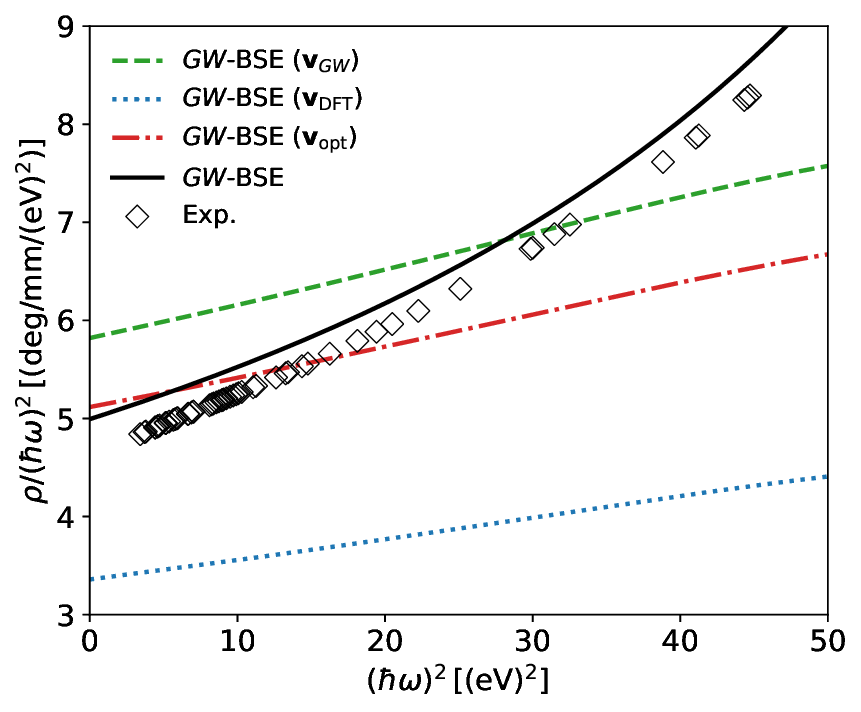}
\caption{\label{fig2} Energy-renormalized Optical rotatory dispersion of $\alpha$-quartz. Experimental points are taken from Ref.~\cite{lowry1964}. The solid curve is obtained from the SOXS formulation.}
\end{figure}

We compare two formulations of excitonic optical activity of $\alpha$-quartz by plotting the energy-renormalized optical rotation $\rho/(\hbar \omega)^2$ as a function of $(\hbar \omega)^2$, which more clearly reveals the frequency dependence (Fig.~\ref{fig2}). For the envelope modulation approach, the different treatments of the velocity matrix elements yield very similar dispersions, all of which deviate from experiment. In contrast, the SOXS formulation reproduces the experimental frequency dependence. This highlights the critical importance of treating the velocity operator consistently within the exciton oscillator strength.

In summary, we have developed a many-body theory of excitonic optical activity in solids within the \textit{GW}–BSE framework. Applied to the prototypical chiral crystal $\alpha$-quartz, our approach yields optical rotations in excellent agreement with experiment. By providing a predictive, \textit{ab initio} description of optical activity, this work advances fundamental understanding of chiral light–matter interaction in solids and opens a route to rational design of chiroptoelectronic materials.

\begin{acknowledgments}
This work was supported as part of the Center for Hybrid Organic Inorganic Semiconductors for Energy (CHOISE) an Energy Frontier Research Center funded by the Office of Basic Energy Sciences, Office of Science within the U.S. Department of Energy. The calculations were performed using computational resources of the National Energy Research Scientific Computing Center (NERSC), a US Department of Energy Office of Science User Facility located at Lawrence Berkeley National Laboratory, operated under contract DE-AC02-05CH11231 using NERSC award BES-ERCAP0023945, and resources sponsored by the Department of Energy's Office of Energy Efficiency and Renewable Energy and located at the National Laboratory of the Rockies.  
\end{acknowledgments}

\email{xiaoming.wang}
\bibliography{refs}
\bibliographystyle{apsrev4-2}

\end{document}



\title{Supplementary Material for ``\textit{Ab initio} \textit{GW}-BSE theory of optical activity in $\alpha$-quartz''}

\author{Xiaoming Wang}
\email{xiaoming.wang@utoledo.edu}
\affiliation{%
 Department of Physics and Astronomy, Wright Center for Photovoltaics Innovation and Commercialization, The University of Toledo, Toledo, Ohio 43606, USA
}%

\author{Yanfa Yan}%
\email{yanfa.yan@utoledo.edu}
\affiliation{%
 Department of Physics and Astronomy, Wright Center for Photovoltaics Innovation and Commercialization, The University of Toledo, Toledo, Ohio 43606, USA
}%


\maketitle

\section{Derivation of Eq. (15)}
We focus on the first order term of Eq. (15). Inserting Eqs. (13) and (14) in $\langle \Psi_\lambda (\mathbf{q})|e^{i\mathbf{q \cdot r}}|\Psi_\mu \rangle $ and writing $A_{cv\mathbf{k}}^\lambda (\mathbf{q})$ as $A_{c\mathbf{k+q}, v\mathbf{k}}^\lambda$, leads to
\begin{equation}
    \sum_{cv\mathbf{k}}(A_{c\mathbf{k+q},v\mathbf{k}}^{\lambda })^* \sum_{c'v'\mathbf{k}'} A_{c'\mathbf{k}',v'\mathbf{k'}}^{\mu} \sum_{nn'\mathbf{k''}} \langle u_{n\mathbf{k''+q}}|u_{n'\mathbf{k''}} \rangle \langle 0| c_{v\mathbf{k}}^\dagger c_{c\mathbf{k+q}} c_{n\mathbf{k'' + q}}^\dagger c_{n' \mathbf{k''}} c_{c'\mathbf{k}'}^\dagger c_{v'\mathbf{k}'} |0 \rangle ,
\end{equation}
where the operator product can be simplified with Wick's theorem
\begin{equation}
    \begin{split}
    c_{v\mathbf{k}}^\dagger c_{c\mathbf{k+q}} c_{n\mathbf{k'' + q}}^\dagger c_{n' \mathbf{k''}} c_{c'\mathbf{k}'}^\dagger c_{v'\mathbf{k}'} &= \, \bcancel{ :c_{v\mathbf{k}}^\dagger c_{c\mathbf{k+q}} c_{n\mathbf{k'' + q}}^\dagger c_{n' \mathbf{k''}} c_{c'\mathbf{k}'}^\dagger c_{v'\mathbf{k}'}: } 
    + \, \bcancel{\delta_{cn}\delta_{\mathbf{kk''}}:c_{v\mathbf{k}}^\dagger c_{n' \mathbf{k''}} c_{c'\mathbf{k}'}^\dagger c_{v'\mathbf{k}'}:} \\
    &+ \bcancel{\delta_{n'c'}\delta_{\mathbf{k''k'}}:c_{v\mathbf{k}}^\dagger c_{c\mathbf{k+q}} c_{n\mathbf{k'' + q}}^\dagger c_{v'\mathbf{k}'}:} 
    + \, \delta_{cc'}\delta_{\mathbf{k+q,k'}}:c_{v\mathbf{k}}^\dagger c_{n\mathbf{k'' + q}}^\dagger c_{n' \mathbf{k''}}  c_{v'\mathbf{k}'}: \\
    &+ \delta_{cn}\delta_{\mathbf{kk''}}\delta_{n'c'}\delta_{\mathbf{k''k'}}:c_{v\mathbf{k}}^\dagger  c_{v'\mathbf{k}'}: \, \\
    &= -\delta_{cc'}\delta_{\mathbf{k+q,k'}}\delta_{vn'}\delta_{\mathbf{kk''}}\delta_{nv'} + \, \delta_{cn}\delta_{\mathbf{kk''}}\delta_{n'c'}\delta_{\mathbf{k'',k'}}\delta_{vv'} \, .
\end{split}
\end{equation}
Inserting Eq. (S2) in Eq. (S1), we have
\begin{equation}
    -\sum_{cvv'\mathbf{k'}} (A_{c\mathbf{k'},v\mathbf{k'-q}}^\lambda)^* A_{c\mathbf{k'},v'\mathbf{k'}}^\mu \langle u_{v'\mathbf{k'}} | u_{v\mathbf{k'-q}} \rangle + \sum_{cc'v\mathbf{k}} (A_{c\mathbf{k+q},v\mathbf{k}}^\lambda)^* A_{c'\mathbf{k},v\mathbf{k}}^\mu \langle u_{c\mathbf{k+q}} | u_{c'\mathbf{k}} \rangle .
\end{equation}
Now, we expand Eq. (S3) to first order of $\mathbf{q}$. Firstly, expand the Bloch overlap terms (changing dummy index $\mathbf{k}'$ to $\mathbf{k}$)
\begin{equation}
    \begin{split}
        \langle u_{v'\mathbf{k}} | u_{v\mathbf{k-q}} \rangle &= - \mathbf{q} \cdot \langle u_{v'\mathbf{k}} | \partial_{\mathbf{k}} u_{v\mathbf{k}} \rangle = i\mathbf{q} \cdot \bm{\mathcal{A}}_{v'v\mathbf{k}} \, , \\
        \langle u_{c\mathbf{k+q}} | u_{c'\mathbf{k}} \rangle &=  \mathbf{q} \cdot \langle \partial_{\mathbf{k}} u_{c\mathbf{k}} | u_{c'\mathbf{k}} \rangle = i\mathbf{q} \cdot \bm{\mathcal{A}}_{cc'\mathbf{k}} \, .
    \end{split}
\end{equation}
Including zeroth-order exciton envelopes in Eq. (S4), yields
\begin{equation}
    i\mathbf{q} \cdot \left[ \sum_{cc'v\mathbf{k}} (A_{cv\mathbf{k}}^\lambda)^* A_{c'v\mathbf{k}}^\mu \bm{\mathcal{A}}_{cc'\mathbf{k}} - \sum_{cvv'\mathbf{k}} (A_{cv\mathbf{k}}^\lambda)^* A_{cv'\mathbf{k}}^\mu \bm{\mathcal{A}}_{v'v\mathbf{k}} \right] \, .
\end{equation}
If we take zeroth-order of the Bloch overlaps, Eq. (S3) survives only for $v'=v$ and $c=c'$, and becomes
\begin{equation}
    \sum_{cv\mathbf{k}} \left[ (A_{c\mathbf{k+q},v\mathbf{k}}^\lambda)^* - (A_{c\mathbf{k},v\mathbf{k-q}}^\lambda)^*\right] A_{c\mathbf{k},v\mathbf{k}}^\mu  = \mathbf{q} \cdot \sum_{cv\mathbf{k}} \partial_\mathbf{k} (A_{c\mathbf{k},v\mathbf{k}}^\lambda)^* A_{c\mathbf{k},v\mathbf{k}}^\mu \\
     = i\mathbf{q} \cdot \sum_{cv\mathbf{k}} (A_{cv\mathbf{k}}^\lambda)^\ast i\partial_\mathbf{k} A_{cv\mathbf{k}}^\mu \, .
\end{equation}
We see that Eqs. (S5) and (S6) (removing prefactor $i\mathbf{q}$) are exactly the same as Eqs. (16) and (17) with different grouping of terms. 

\section{Derivation of inter-exciton transition dipole moment}
To evaluate $\bm{\mathcal{R}}_{\lambda \mu} = \langle \Psi_\lambda |\mathbf{r}|\Psi_\mu \rangle$, we use
\begin{equation}
    |\Psi_\lambda \rangle = \sum_{cv\mathbf{k}} A_{cv\mathbf{k}}^\lambda c_{c\mathbf{k}}^\dagger c_{v\mathbf{k}} |0\rangle , \quad |\Psi_\mu \rangle = \sum_{c'v'\mathbf{k}'} A_{c'v'\mathbf{k}'}^\lambda c_{c'\mathbf{k}'}^\dagger c_{v'\mathbf{k}'} |0\rangle ,
\end{equation}
and the position operator
\begin{equation}
    \mathbf{r} = \sum_{nm\mathbf{k}'' \mathbf{k}'''} \langle u_{n\mathbf{k}''} |\mathbf{r} | u_{m\mathbf{k}'''} \rangle c_{n\mathbf{k}''}^\dagger c_{m\mathbf{k}'''}.
\end{equation}
Note that the matrix elements of the position operator should be understood as
\begin{equation}
    \mathbf{r}_{nm\mathbf{k}'' \mathbf{k}'''} = (1-\delta_{nm})\delta_{\mathbf{k}'' \mathbf{k}'''} \bm{\mathcal{A}}_{nm\mathbf{k}''} + \delta_{nm}(\delta_{\mathbf{k}'' \mathbf{k}'''} \bm{\mathcal{A}_{nn\mathbf{k}''}} + i\bm{\nabla}_{\mathbf{k}''} \delta_{\mathbf{k}'' \mathbf{k}'''}) \, ,
\end{equation}
where the first and second terms correspond to the \emph{interband} and \emph{intraband} contributions, respectively. Inserting Eqs. (S7) and (S8), we have 
\begin{equation}
    \bm{\mathcal{R}}_{\lambda \mu} = \langle 0 |\sum_{cv\mathbf{k}} (A_{cv\mathbf{k}}^{\lambda})^\ast c_{v\mathbf{k}}^\dagger c_{c\mathbf{k}} \sum_{nm\mathbf{k'' k'''}} \mathbf{r}_{nm\mathbf{k'' k'''}} c_{n\mathbf{k''}}^\dagger c_{m\mathbf{k'''}} \sum_{c' v' \mathbf{k}'} A_{c' v' \mathbf{k}'}^\mu c_{c' \mathbf{k}'}^\dagger c_{v' \mathbf{k}'} |0\rangle,
\end{equation}
where the operator product can be simplified with Wick's theorem
\begin{equation}
    \begin{split}
        c_{v\mathbf{k}}^\dagger c_{c\mathbf{k}} c_{n\mathbf{k}''}^\dagger c_{m\mathbf{k}'''} c_{c' \mathbf{k}'}^\dagger c_{v' \mathbf{k}'} =& \bcancel{\mathord{:} c_{v\mathbf{k}}^\dagger c_{c\mathbf{k}} c_{n\mathbf{k}''}^\dagger c_{m\mathbf{k}'''} c_{c' \mathbf{k}'}^\dagger c_{v' \mathbf{k}'} \mathord{:} }
        + \bcancel{\delta_{cn} \delta_{\mathbf{kk''}} \mathord{:} c_{v\mathbf{k}}^\dagger  c_{m\mathbf{k}'''} c_{c' \mathbf{k}'}^\dagger c_{v' \mathbf{k}'} \mathord{:}} \\
        +& \bcancel{\delta_{mc'} \delta_{\mathbf{k''' k'}} \mathord{:} c_{v\mathbf{k}}^\dagger c_{c\mathbf{k}} c_{n\mathbf{k''}}^\dagger  c_{v' \mathbf{k}'} \mathord{:} } 
        + \delta_{cc'} \delta_{\mathbf{kk}'} \mathord{:} c_{v\mathbf{k}}^\dagger c_{n\mathbf{k''}}^\dagger c_{m\mathbf{k'''}} c_{v' \mathbf{k}'} \mathord{:} \\
        +& \delta_{cn} \delta_{\mathbf{kk''}} \delta_{mc'} \delta_{\mathbf{k''' k'}} \mathord{:} c_{v\mathbf{k}}^\dagger  c_{v' \mathbf{k}'} \mathord{:}
    \end{split}
\end{equation}
The last term can be nonzero only when $v = v'$ and $\mathbf{k} = \mathbf{k'}$. Inserting this term in Eq. (S10), we have the interband contribution
\begin{equation}
    \sum_{cv\mathbf{k}} \sum_{c' v' \mathbf{k}'} \sum_{nm \mathbf{k''k'''}} (A_{cv\mathbf{k}}^{\lambda})^\ast A_{c' v' \mathbf{k} '}^\mu \mathbf{r}_{nm\mathbf{k''k'''}} \delta_{cn} \delta_{\mathbf{kk''}} \delta_{mc'} \delta_{\mathbf{k'''k'}} \delta_{vv'} \delta_{\mathbf{kk}'} 
    = \sum_{cv\mathbf{k}, c' \neq c} (A_{cv\mathbf{k}}^{\lambda})^\ast A_{c' v \mathbf{k}}^\mu \bm{\mathcal{A}}_{cc' \mathbf{k}} \, .
\end{equation}
For the intraband part, we should evaluate the derivative before assigning $\mathbf{k}=\mathbf{k}'$, yielding
\begin{equation}
    \sum_{cv\mathbf{k}} (A_{cv\mathbf{k}}^{\lambda})^\ast A_{c v \mathbf{k}}^\mu \bm{\mathcal{A}}_{cc\mathbf{k}} + i (A_{cv\mathbf{k}}^{\lambda})^\ast \partial_\mathbf{k} A_{c v \mathbf{k}}^\mu \, .
\end{equation}
There are two possible manipulations for the second to last term of Eq. (S11) to be nonzero: $D1 = \delta_{cc'} \delta_{\mathbf{kk}'} \delta_{nm} \delta_{\mathbf{k''k'''}} \delta_{vv'}$ and $D2 = -\delta_{cc'} \delta_{\mathbf{kk}'} \delta_{nv'} \delta_{\mathbf{k''k'}} \delta_{vm} \delta_{\mathbf{kk'''}}$. Inserting D1 in Eq. (S10), we have
\begin{equation}
    \sum_{cv\mathbf{k}} \sum_{c' v' \mathbf{k}'}  (A_{cv\mathbf{k}}^{\lambda})^\ast A_{c' v' \mathbf{k} '}^\mu \delta_{cc'} \delta_{vv'} \delta_{\mathbf{kk}'} \sum_{nm \mathbf{k''k'''}} \mathbf{r}_{nm\mathbf{k''k'''}} \delta_{nm} \delta_{\mathbf{k''k'''}} = 0,
\end{equation}
thanks to the orthogonality $\sum_{cv\mathbf{k}} (A_{cv\mathbf{k}}^{\lambda})^\ast A_{cv\mathbf{k}}^\mu =0 $. Inserting D2 in Eq. (S10), we obtain
\begin{equation}
    -\sum_{cv\mathbf{k}, v' \neq v} (A_{cv\mathbf{k}}^{\lambda})^\ast  A_{cv' \mathbf{k}}^\mu \bm{\mathcal{A}}_{v' v \mathbf{k}} \, ,
\end{equation}
for the interband contribution, and the intraband component
\begin{equation}
    -\sum_{cv\mathbf{k}} (A_{cv\mathbf{k}}^{\lambda})^\ast A_{c v \mathbf{k}}^\mu\bm{\mathcal{A}}_{vv\mathbf{k}} + \bcancel{i [(A_{cv\mathbf{k}}^{\lambda})^\ast \partial_\mathbf{k} A_{c v \mathbf{k}}^\mu + \partial_\mathbf{k} (A_{cv\mathbf{k}}^{\lambda})^\ast A_{c v \mathbf{k}}^\mu]} .
\end{equation}
Combining Eqs. (S12--S13) and (S15--S16), we arrive at the Eqs. (16) and (17) of $\bm{\mathcal{R}}_{\lambda \mu}$.

\section{convergence}

\begin{figure}
\includegraphics[width=0.6\linewidth]{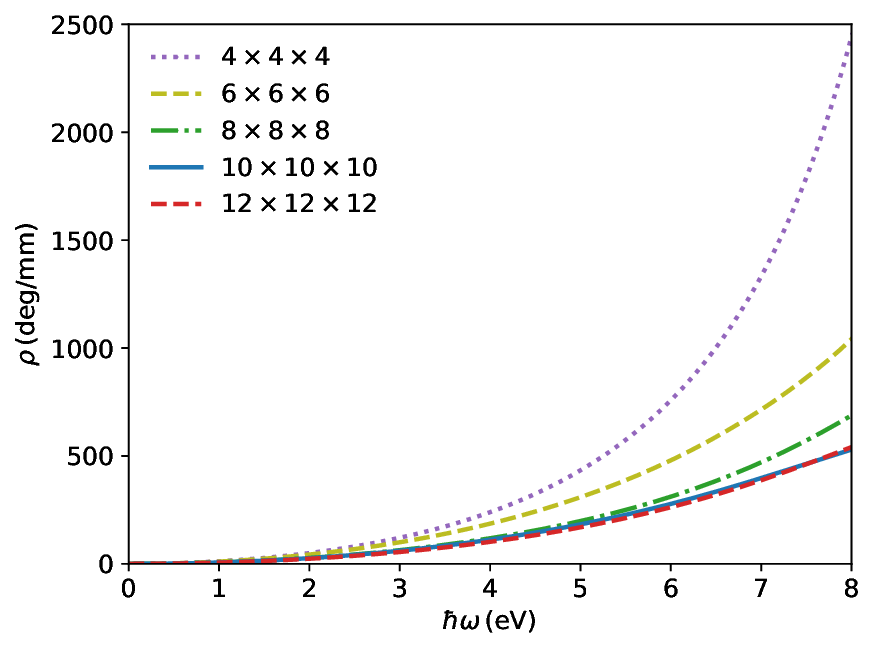}
\caption{\label{figs1} Convergence of the optical rotation of $\alpha$-quartz on k-point mesh.}
\end{figure}

\begin{figure}
\includegraphics[width=\linewidth]{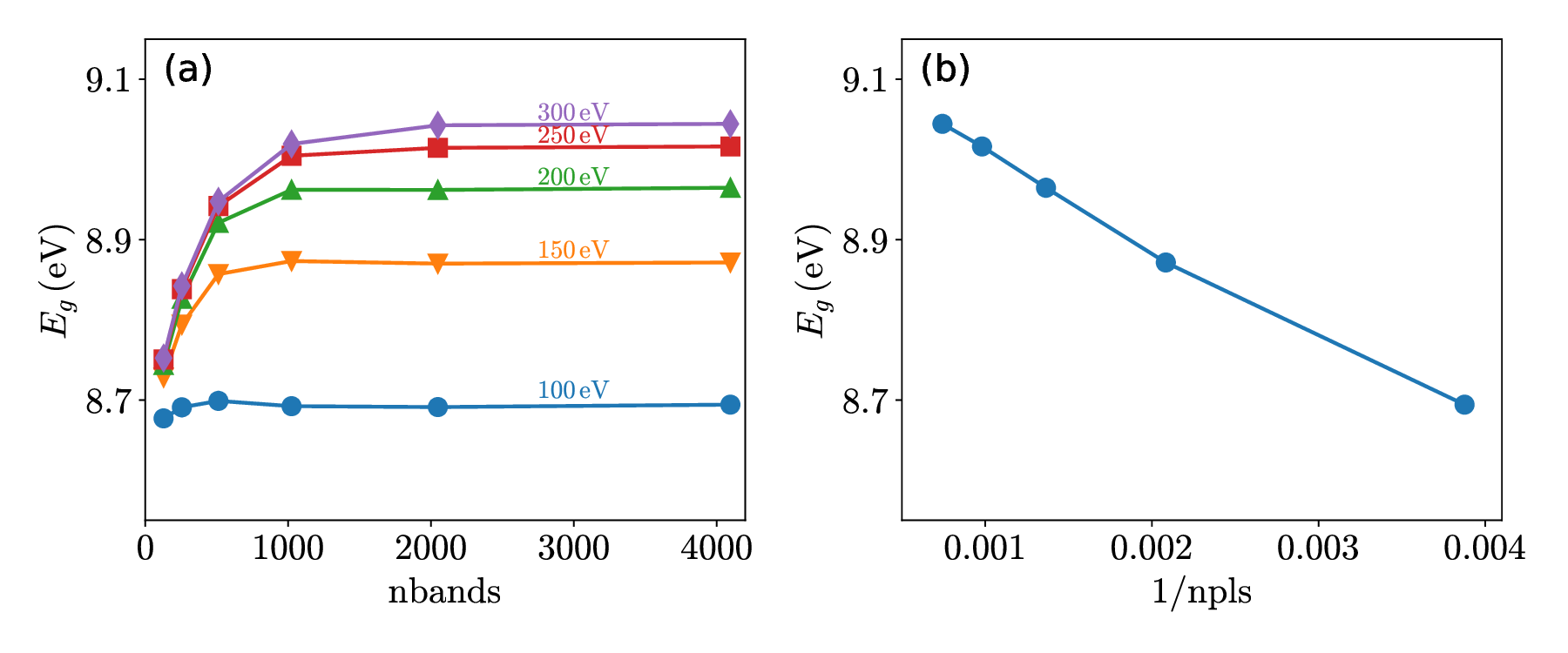}
\caption{\label{figs1} (a) Convergence of number of bands at different dielectric cutoff energies (ENCUTGW) for \textit{GW} calculations. (b) Basis set (number of plane waves) extrapolation for \textit{GW} band gap.}
\end{figure} 
